\documentclass[boxit]{JAC2003}

\usepackage{graphicx}
\usepackage{booktabs}
\usepackage{verbatim}
\usepackage{subcaption}
\begin{document}

\title{Status of the SPP RFQ project}

\author{G. Turemen\thanks{gorkem.turemen@ankara.edu.tr}, B. Yasatekin, Ankara University, Ankara, Turkey\\ 
H.Yildiz, Istanbul University, Istanbul, Turkey\\ 
A. Alacakir, TAEK-SANAEM, Ankara, Turkey\\ 
G. Unel, UCI, Irvine, California, USA\\
}

\maketitle

\begin{abstract}
 The SPP project at TAEK will use a 352.2 MHz 4-vane Radio Frequency Quadrupole (RFQ) to accelerate $H+$ ions from 20 keV to 1.5 MeV. With the design already complete, the project is at the test production phase. To this effect, a so called ``cold model'' of 50 cm length has been produced to validate the design approach, to perform the low power RF tests and to evaluate possible production errors. This study will report on the current status of the low energy beam transport line (LEBT) and RFQ cavity of the SPP project. It will also discuss the design and manufacturing of the RF power supply and its transmission line. In addition, the test results from some of the LEBT components will be shown and the final RFQ design will be shared. 
\end{abstract}

\section{Introduction}

SANAEM
\footnote{SANAEM is one of the two research centers of TAEK, Turkish Atomic
Energy Agency.
} Prometheus Project aims to construct a proton beamline including
an ion source, a Proof Of Principle (POP) RFQ and the appropriate
diagnostic stations \cite{ibic13}. The POP machine has the humble
requirements of accelerating 1 mA of beam current to an energy of 1.5
MeV. The entire system is to be designed and built in Turkey, with
the main goal of training young accelerator physicists and RF engineers
on the job and to involve the local industry in accelerator component
construction. The project design and construction is planned to last
for 3 years, and it is to start operation by the end of 2015.

\section{Status of the SPP beamline}

The SPP project will construct an ion source, a low energy beam transport
(LEBT) section including diagnostics at keV level, an RFQ receiving
power from an RF Power Supply Unit (PSU) via a transfer line, a MeV level diagnostics
and finally a beam dump. The RFQ operating frequency will be 352.2
MHz. The beamline will be sitting on a locally designed and built
pedestal. The cooling for the RF components, ion source, magnets and the cavity will be
provided by a 40 kW closed loop chiller. The chiller and the pedestal supporting the ion source and two solenoid magnets are installed in the experimental area.

\subsection{Ion Source, LEBT and Diagnostic Stations}

\begin{figure}[!htb]
   \centering
   \includegraphics[width=0.47\textwidth]{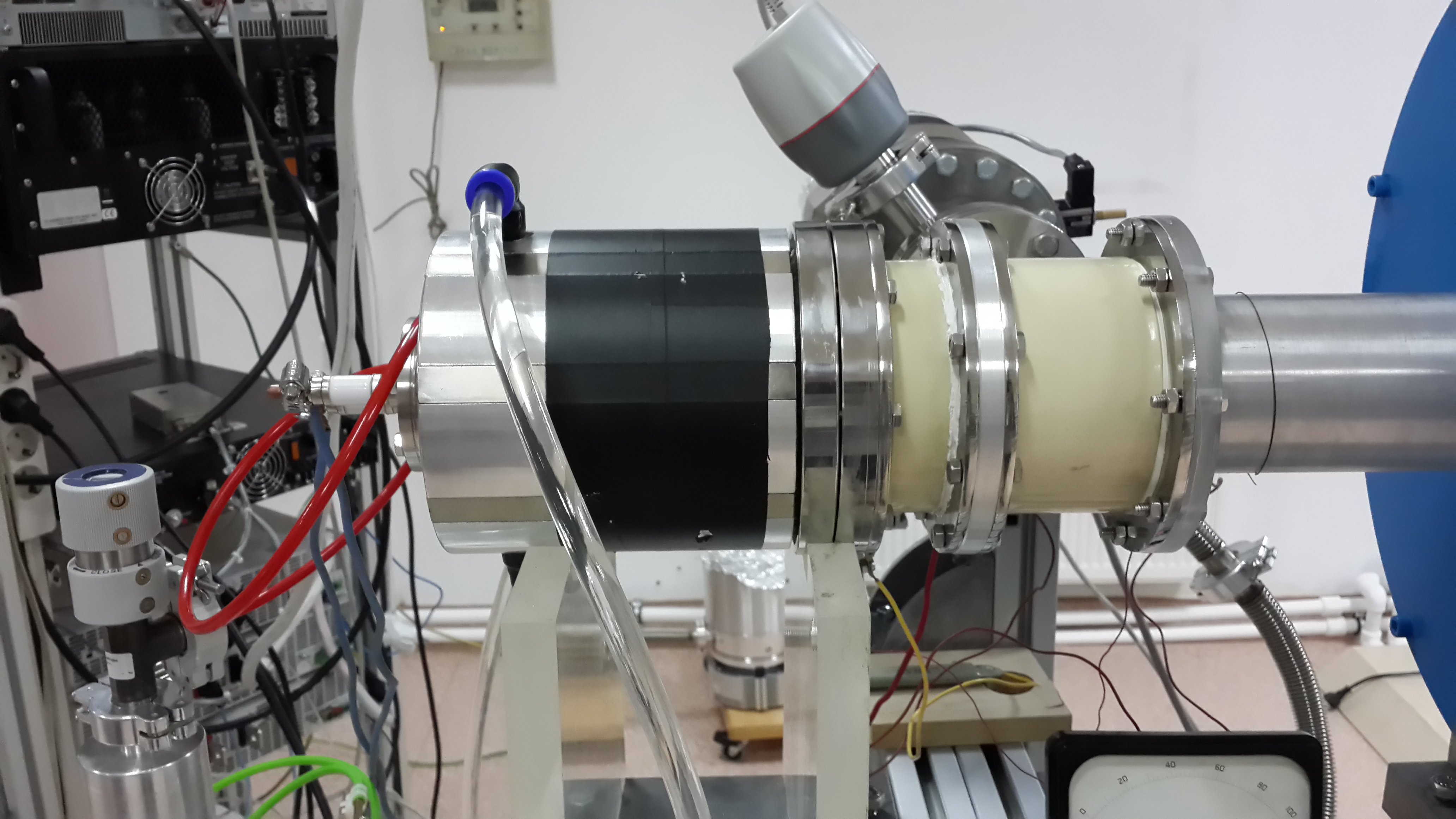}
   \caption{Ion source during the tests.}
   \label{ion}
\end{figure}

The ion source and LEBT design was discussed extensively elsewhere
\cite{ipac-design}. Here we mostly report on the ongoing construction
and installation processes. The ion source, as well as the first (Sol-M) of the
two LEBT solenoids (Sol-M and Sol-L) are installed together with the vacuum and cooling
connections as it can be seen on Fig.\ref{ion}.
While waiting the delivery of the ``measurements box'', current
and stability tests are performed with a simple Faraday cup (FC) for two different ion source configurations (RF antenna and DC filament). Also the stability (magnetic field) tests are done for LEBT solenoids. Measured time required for Sol-M and Sol-L to reach steady state are approximately 111 and 105 minutes respectively.

\begin{figure}[!htb]
   \centering
   \includegraphics[width=0.47\textwidth]{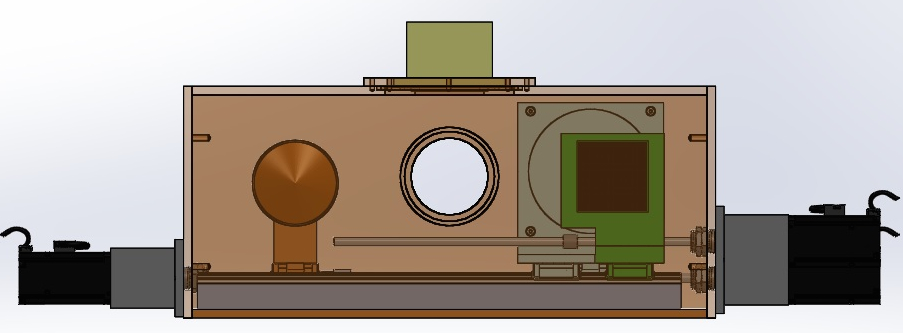}
   \caption{Measurement box design.}
   \label{box}
\end{figure}

The measurements box will be located between the two solenoids and
it will contain a FC, a pepper-pot filter for emittance measurements
and a scintillator screen. All components will be motor controlled
such that they can be pushed in and out of the beam independently.
Fig.\ref{box} shows the measurement box and
its components in parking position.

\subsection{SPP RFQ Design Studies}

Two different RFQs of various properties, designed with LIDOS
\cite{lidos} and cross checked with DEMIRCI \cite{demirci} and TOUTATIS
\cite{toutatis} software packages, are shown in Table \ref{Different-RFQ-designs}.
The property of primary importance will be the RFQ length as by the
time of this note, no brazing furnace of appropriate length is available
in Turkey. Therefore while waiting for such a furnace, it was decided
to follow a two-step approach: step-1) construction of an Al
+ Cu coated RFQ, from a single Al piece machined on a 120 cm CNC to
help fulfill the project goals; step-2) after securing of a proper
brazing furnace, construction of a two section OFE Cu RFQ. The first
RFQ is planned to be a modified version of design-A, with improved
transmission and the second one will be based on design-B with improved
output energy.

\begin{table}[hbt]
   \centering
   \caption{Two Different RFQ designs and their main properties}
   \begin{tabular}{cccc}
       \toprule
       \textbf{Design} & \textbf{L (cm)}  & \textbf{E$_{out}$ (MeV)} & \textbf{T$_{tot}$/T$_{acc}$ (\%)} \\
       \midrule
          A & 119.0 & 1.3 & 99.0/94.1      \\
          B & 164.6 & 1.5 & 99.7/96.2      \\
       \bottomrule
   \end{tabular}
   \label{Different-RFQ-designs}
\end{table}

\subsubsection{Electromagnetic Considerations}~\\

The 2D electromagnetic design is modified after discussions on cooling and machining of the RFQ with CNC experts. Comparisons of quality factor and power dissipation for the current and the previous design\cite{ibic13} is shown in Table \ref{2dEM}. To ease RF tuning (considering the tuning ranges of the tuners) of the cavity the resonant frequency is set to 1.2 MHz below the actual frequency.

\begin{table}[hbt]
   \centering
   \caption{Parameters of the 2D electromagnetic design of SANAEM-RFQ}
   \begin{tabular}{cccc}
       \toprule
       \textbf{Design} & \textbf{F (MHz)}  & \textbf{Q (\#)} & \textbf{P$_{SF}$ (W/cm)} \\
       \midrule
          Prior & 352.2 & 10342 & 126.0      \\
          Current & 351.0 & 10499 & 121.5      \\
       \bottomrule
   \end{tabular}
   \label{2dEM}
\end{table}

Therefore, the vanes were thickened to 14 mm at 35 mm from the beam axis to provide sufficient space for cooling channels and also a vane base width increased by approximately 5 mm to reduce the mechanical vibrations.

\subsection{Cold Model Tests}

\begin{figure}[!htb]
   \centering
   \includegraphics[width=0.95\columnwidth]{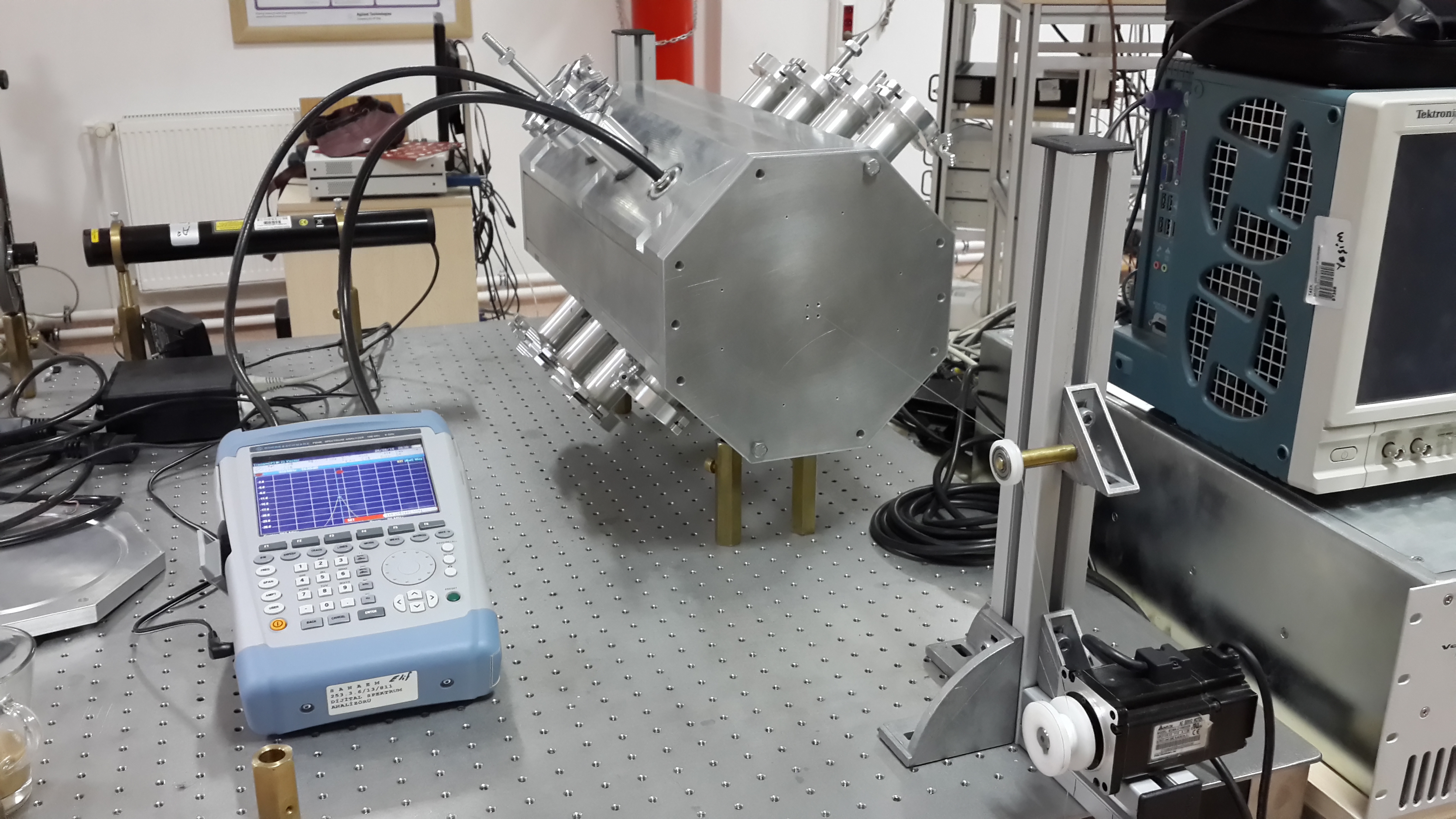}
   \caption{Constructed SANAEM-RFQ ``cold model'' while bead pull measurements.}
   \label{coldmodel}
\end{figure}

50 cm length Al cold model (Fig. \ref{coldmodel}) is produced to perform the low power RF tests, to gain experience of production of RFQ vanes and evaluate possible production errors. 

\begin{figure}[!htb]
   \centering
\includegraphics[width=0.95\columnwidth]{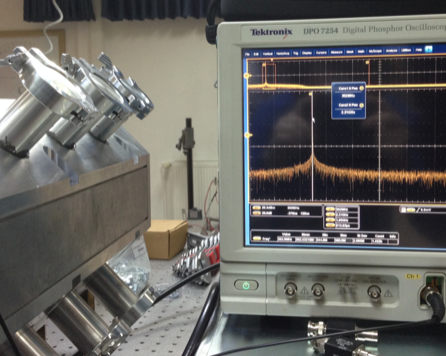}
\caption{Resonant frequency tests of the cold model.}
   \label{resonant}
\end{figure}

Desired cavity resonant frequency, 352.2 MHz, was obtained at the RF tests (Fig. \ref{resonant}). Also the effect of the tuners on resonant frequency is studied (tuning range was about -100 and +400 kHz). The bead-pull tests are ongoing to tune and obtain required field flatness inside the cavity.

\subsection{RF Power and Transmission Line}

\subsubsection{PSU}~\\

\begin{figure}[!htb]
   \centering
\includegraphics[width=0.95\columnwidth]{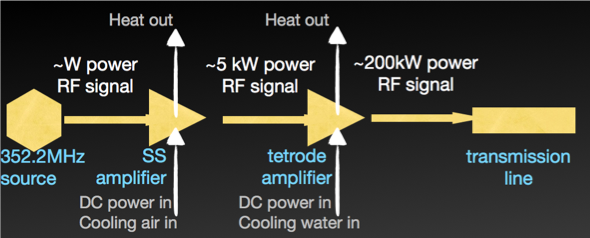}
   \caption{The two stage RF power amplification scheme for SPP PSU.}
   \label{RF-psu}
\end{figure}

The power requirements of the SPP RFQ have been calculated to be about
170 kW, including the ohmic loss and beam loading. Adding in the losses
on the transmission line and the overheads, the expected RF power
needed for the beamline is about 200 kW. A power supply to match these
requirements is being built by a private company \cite{eprom} under guidance from
the SPP team. The amplification will be a hybrid one and will be achieved
in two stages consisting of solid state (SS) and tetrode amplifiers.
The schematic view of this hybrid PSU is shown in Fig. \ref{RF-psu}.
The SS section is partly constructed by using BLF578 integrated circuit
and is able to provide 2kW RF power. The selected tetrode is the water
cooled TH595 which has an amplification power of 15dB and a duty factor
of about 3\% \cite{TH595}. The PSU output is envisaged to be of coaxial
type with 3 1/8 inch diameter.

\subsubsection{The Circulator and the Transmission line}~\\

\begin{figure}[!htb]
   \centering
\includegraphics[width=0.92\columnwidth]{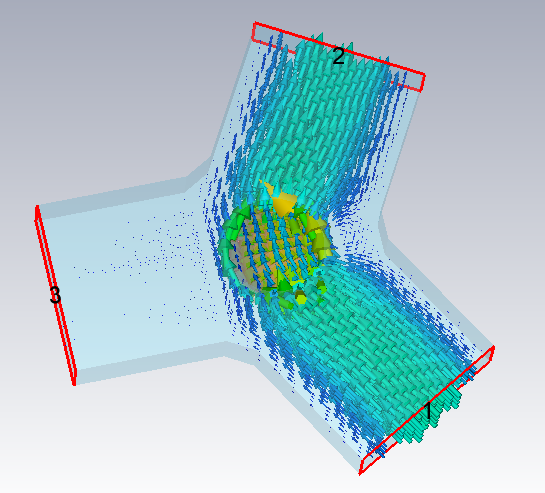}
   \caption{Current circulator design.}
   \label{circ}
\end{figure}

After considering stripline, full height (FH) and half height (HH)
models, the preliminary circulator design is made with the last option.
The PSU coaxial line will initially be converted to a HH waveguide
which will subsequently enter the circulator port 1. The circulator
itself is selected to operate in above resonance mode and a set of
commercially available Yttrium based garnets are scanned to find the
optimum material, geometry and bias field to apply. So far, the results
show that over 90\% power transmission can be achieved with about
1\% or less return and isolation losses. Further optimization of the
circulator design and the procurement of the ferrite material are
ongoing (Fig. \ref{circ}). 

\begin{figure}[!htb]
   \centering
\includegraphics[width=0.92\columnwidth]{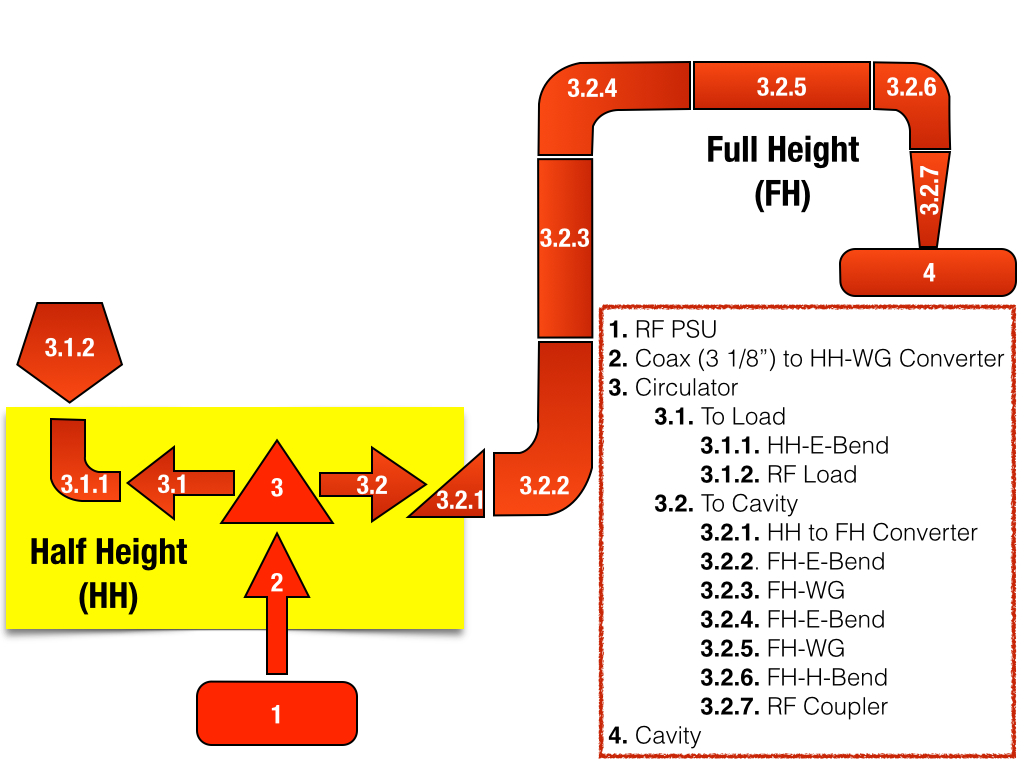}
   \caption{RF transfer line design.}
   \label{RF-transferline}
\end{figure}

The circulator output towards the RFQ will be converted to FH with
a dedicated section right after the output and the power will be transferred
to the RF power in FH to benefit from the already existing components.
A schematic view of the RF transmission line can be seen in Fig. \ref{RF-transferline}.
In the schema, the components with yellow background are HH and the
rest is FH WR2300 rectangular waveguides.

\section{Conclusions}

This paper summarizes the status of the SPP beamline and the RF components, including the prototype tests. The ion source and the LEBT are built and the beam tests are ongoing. There are a few different RFQ designs under consideration being evaluated for manufacturing and assembly options. The project aims to obtain the first MeV range protons by the end of 2015. For the MeV level diagnostics, current measurement by an ACCT and energy measurement by a spectrometer is planned. These components will also be home designed and built. The ACCT prototype is ready and the construction of the wire chamber for the spectrometer is ongoing.

\section{Acknowledgements}
This project is funded by TAEK with a project code A4.H4.P1. The authors are grateful to S. Ogur and S. Oz for useful comments and fruitful discussions. The authors would like to thank Z. Sali for increasing their expertise in electromagnetic simulations and ditto M. Celik for technical drawings.



\begin{thebibliography}{99} 

\bibitem{ibic13}
G. Turemen, B. Yasatekin, O. Mete, M. Celik, Z. Sali, Y. Akgun, A. Alacakir, S. Bolukdemir, E. Durukan, H. Karadeniz, E. Recepoglu, E. Cavlan, G. Unel and S. Erhan,
  ``Project PROMETHEUS: Design and Construction of a Radio Frequency Quadrupole at TAEK,''
  IBIC'13, Oxford, September 2013, WEPC02, p.~656.\\
http://arxiv.org/pdf/1310.0790v1.pdf

\bibitem{ipac-design}
G. Turemen, Z. Sali, B. Yasatekin, V. Yildiz, M. Celik, A. Alacakir, G. Unel, O. Mete,
``SPP Beamline Design and Beam Dynamics'', IPAC'14, Dresden, June 2014, THPME050, p. 3338.
http://arxiv.org/pdf/1406.3066.pdf

\bibitem{lidos}
``LIDOS.RFQ.DESIGNER''
Version 1.3\\
http://www.ghga.com/accelsoft

\bibitem{demirci}
B.Yasatekin, G.Turemen, G.Unel,
``A Graphical Approach to Radio Frequency Quadrupole Design''\\
http://arxiv.org/pdf/1401.2196.pdf

\bibitem{toutatis}
R. Duperrier,
``TOUTATIS: A radio frequency quadrupole code''
Phys. Rev. Vol.3, 124201, 2000.


\bibitem{eprom}
EPROM Elektronik, Ankara, Turkey.\\
http://www.epromvector.com/about.html





\bibitem{TH595}
https://www.thalesgroup.com/sites/default/files/asset/document/PowerGrid


\end{thebibliography}
\end{document}